\documentclass[aps,twocolumn,superscriptaddress,floatfix,showpacs,preprintnumbers,amsmath,amssymb]{revtex4-1}

\usepackage[dvips]{graphicx}
\usepackage{color}
\usepackage{dsfont}
\usepackage{bm}
\usepackage{amsmath}
\usepackage{amsfonts}
\usepackage{psfrag}

\newcommand{\dD}{\textrm{d}} 

\newcommand{\I}{\textrm{i}}
\newcommand{\E}{\textrm{e}}

\newcommand{\mwe}[1]{\left\langle\,#1\right\rangle_\eta}

\newcommand{\mw}[1]{\left\langle\, #1\right\rangle}
\newcommand{\cumu}[1]{\left\langle\left\langle\, #1\right\rangle\right\rangle} 

\newcommand{\bra}[1]{\ensuremath{\left\langle #1\right|}}
\newcommand{\ket}[1]{\ensuremath{\left|#1\right\rangle}}

\begin{document}

\title{Linear dynamics subject to thermal fluctuations and non-Gaussian noise:\\
 From classical to quantum}

\author{M. K\"opke}
\affiliation{Max Planck Institute for Solid State Research, Heisenbergstrasse 1, D-70569 Stuttgart, Germany}
\affiliation{Institut f\"ur Theoretische Physik, Universit\"at Ulm, Albert-Einstein-Allee 11, 89069 Ulm, Germany}
\author{J. Ankerhold}
\affiliation{Institut f\"ur Theoretische Physik, Universit\"at Ulm, Albert-Einstein-Allee 11, 89069 Ulm, Germany}

\date{\today}

\begin{abstract}
The dynamics of a linear system embedded in a heat bath environment and subject to white non-Gaussian noise is studied. Classical higher order cumulants in coordinate space are derived for Poissonian noise and their impact on the dynamics and on asymptotic steady state distributions is analyzed.
In the quantum regime non-Gaussian properties are present in the reduced density in coordinate representation which  in energy representation exist on a transient time scale only due to symmetry. Within an exactly solvable model our results provide insight into mechanisms of linear detectors as sensors for non-Gaussian noise at high and low temperatures.
\end{abstract}

\pacs{05.40.-a,72.70.+m,05.10.Gg,73.23.-b}

\maketitle

\section{Introduction}

Non-Gaussian noise plays an important role in a variety of fields including nanoscopic solid state structures \cite{blanter,magnoise}, electrical circuits \cite{nazarov}, signal and image processing \cite{signal}, and biological aggregates \cite{bio}. In most cases additional thermal fluctuations are also present due to the interaction of the system of interest with a macroscopic background constituting a heat bath. According to the dissipation fluctuation theorem, the latter one gives also rise to energy exchange such that in absence of non-Gaussian noise the system would equilibrate during its time evolution. The presence of non-Gaussian random forces, typically stationary but out of equilibrium, drive the compound towards a steady state though which may significantly deviate from its thermal equilibrium.

Theoretically, a systematic description for these processes has been developed in the context of charge transfer through mesoscopic conductors. Within the framework of full counting statistics low frequency noise properties for various devices have been determined \cite{nazarov}. For simple devices such as tunnel junctions, predictions have been verified experimentally, cf.~e.g.~\cite{experiments}, but the detection of non-Gaussian noise signatures at finite frequencies is a challenging task. Only very recently has the time series of current noise been monitored down to the $\mu$s-scale in a single-electron  transistor set-up \cite{flindt}.
The same is true for a theoretical understanding which in addition to the noise source must also include the detector degrees of freedom.
In the classical regime, progress has been achieved with Josephson junctions as threshold detectors \cite{jj}. In this set-up the dynamics of the superconducting phase is the relevant coordinate which
is subject to thermal noise and non-Gaussian fluctuations originating from a mesoscopic conductor and acting as an external driving force.
  Its random motion corresponds to that of a fictitious particle in a tilted washboard potential such that the rate of escape from one of the metastable wells encodes information about the non-Gaussian noise \cite{jjtheory}. An extension to the quantum regime including tunneling through the barrier is still elusive, however.

In this paper we consider a much simpler situation, namely, linear dynamics in presence of Poissonian white noise, a problem which allows for a full  analytical treatment. However, the results are by no means trivial and due to the ubiquitous appearance of harmonic systems in all fields of science may be of general interest. In fact, in the context of Josephson junctions this corresponds to the motion of the phase deep inside a well which is known to provide already the correct parameter dependence of the dominating exponential of the escape rate.  There is also another interesting facet of the subject: In the quantum regime the reduced density operator (obtained after integrating out thermal and general non-Gaussian noise) is the relevant object and it is the measurement process which determines its appropriate  representation in Hilbert space. We show that the presence of  non-Gaussian properties in the steady state very sensitively depends on this representation such that e.g.\ for measurements sensitive to energy eigenstates the skewness (third order cumulant) is absent while it can be observed in  coordinate space.  These findings lay on the one hand the basis for implementing a linear detector as a sensor for non-Gaussian noise and on the other hand pave the road  for the treatment of anharmonic systems.

The paper is organized as follows: General results in the classical regime are presented in Secs.~\ref{sec:basics}, \ref{sec:cumulants} which will then be specified to the case of Poissonian noise in Sec.~\ref{sec:classical}. The linear problem can be mapped also to the case of free Brownian motion with Drude damping as illustrated in Sec.~\ref{sec:drude}, a simple model for a tunnel junction in series with an RC-circuit as impedance. The quantum mechanical problem is treated and discussed in Secs.~\ref{sec:qm_pos} and \ref{sec:qm_energy}. Conclusions are given in Sec.~\ref{sec:conclusions}.

\section{Preliminaries}\label{sec:basics}

A linear system with generalized coordinate $q$ embedded in a heat bath and subject to an external force  $F(t)$ can be described by the generalized Langevin equation
\begin{equation}
\ddot{q}(t)+\int_0^t ds\, \gamma(t-s)\, \dot{q}(s)+\omega_0^2 q(t)=\frac{1}{M}[\xi(t)+F(t)]\, .\label{eq:harm-oszi_bewegungsgl}
\end{equation}
Here, $\xi$ is a fluctuating Gaussian force with  $\langle \xi(t)\rangle=0$ which is related to the damping kernel $\gamma(t)$ via a dissipation fluctuation theorem, i.e., $\langle \xi(t) \xi(s)\rangle=(M/\beta)\, \gamma(t-s), \ t>s$, with  inverse temperature $\beta=1/k_{\rm B} T$.

In the sequel, we will mainly concentrate on white Gaussian noise $\gamma(t)=2 \gamma \delta(t)$, but Drude friction $\gamma(t)=\gamma\omega_D\exp(-\omega_D t)$ with a finite bath memory time $1/\omega_D$ will be addressed below (Sec.~\ref{sec:drude}). The external force is taken as a random non-Gaussian variable $F(t)\equiv \eta(t)$ with $\langle \eta(t)\rangle=0$. It is assumed to be statistically independent from the thermal force $\langle\xi(t)\eta(s)\rangle = 0$ and $\delta$-correlated (white non-Gaussian noise). While this certainly simplifies actual experimental situations, it will nevertheless provide insight into the major impact of non-Gaussian noise on the classical and quantum  dynamics of a dissipative oscillator.

Accordingly, we have  solutions of the form  $q(t)=q_0(t)+q_\xi(t) + q_{\eta}(t)$ with the homogeneous part
\begin{equation}
q_0(t)=\E^{-\frac{\gamma}{2}t}\left(c_1\E^{\I\omega t}+c_2\E^{-\I\omega t}\right)\label{eq:harm-oszi_hom-lsg}
\end{equation}
with  $\omega\equiv \sqrt{\omega_0^2-\gamma^2/4}$ and constants $c_1, c_2$ determined by the initial conditions. The force term follows from the response function of the problem
\begin{align}
G(t-t')&=\theta(t-t')\E^{-\gamma(t-t')}\frac{\sin\left(\omega(t-t')\right)}{M \omega}\label{eq:harm-oszi_greensfunktion}
\end{align}
and reads
\begin{equation}\label{inhomo-harmonic}
q_y(t)=\int_0^t ds \, G(t-s) y(s)
\end{equation}
for the thermal noise ($y\equiv\xi$) and the non-Gaussian driving ($y\equiv \eta$), respectively.

\section{Cumulant-Generating Function}\label{sec:cumulants}

We are interested in position correlation functions of the oscillator beyond second order and particularly their steady state properties.
For this purpose, it is convenient to work with cumulant generating functions
\begin{equation}
{\mathcal F}_{q_y}[w]={\rm ln}\langle{\rm e}^{\I\int_0^{\infty}du\,q_{y}(u)w(u)}\rangle_y\ \ \ ; \ \ y= \xi, \eta
\end{equation}
Here, $w(u)$ plays the role of a time dependent counting field. Correlation functions of the full solution $q(t)=q_0(t)+q_\xi(t)+q_\eta(t)$ can then easily be calculated in a straightforward way. Since the noise terms affect only the inhomogeneity of the differential equation, the additive homogeneous solution $q_0$ provides trivial contributions and will be ignored in the sequel. We thus concentrate on the deviations from the deterministic motion. The $n$-th order cumulant then follows from
\begin{eqnarray}
C_{q_y}(t_n, \ldots, t_1)&=&\langle q_y(t_n)\ldots q_y(t_1)\rangle_y\nonumber\\
&=& (-i)^n \left.\frac{\delta^n}{\delta w(t_n)\cdots \delta w(t_1)}{\mathcal F}_{q_y}[w]\right|_{w=0}\, ,
\end{eqnarray}
where $t_n\geq t_{n-1}\cdots \geq t_1$. Cumulants for the momenta $p_y(t)=M \dot{q}_y$ are obtained from taking the corresponding time derivatives.

Now, given the linear structure of the Langevin equation (\ref{eq:harm-oszi_bewegungsgl}), the generating functionals for $q_y$ can be directly inferred  from the generating functionals of the respective noise forces $y$, i.e.,
\begin{equation}
{\mathcal F}_{q_y}[w]={\mathcal F}_{y}[\tilde{w}]\equiv {\rm ln}\langle{\rm e}^{\I\int_0^{\infty}du\, y(u) \tilde{w}(u)}\rangle_y
\end{equation}
with a redefined counting field
\begin{equation}
\tilde{w}(u)=\int_0^\infty ds\  G(s-u) w(s) = \int_u^\infty ds\ G(s-u) w(s)\, .
\end{equation}
Due to the linearity (harmonic system), there is a linear mapping from its cumulants to the characteristics of the noise source via the response function.
Accordingly, for the thermal noise one has $F_\xi[\tilde{w}]=(-M\gamma/\beta) \int_0^\infty du\, \tilde{w}(u)^2$ so that the only non-vanishing cumulant reads
\begin{equation}
C_{q_\xi}(t,t')=(M\gamma/\beta)\int_0^\infty du G(t-u) G(t'-u)\  ;\, t\geq t'\,.
\end{equation}
A compact expression is obtained for the equal-time correlation
\begin{equation}
C_{q_\xi}(t,t)=\frac{1}{M\beta\omega_0^2}-{\rm e}^{-\gamma t} \frac{4\omega_0^2-\gamma^2 \cos(2\omega t)-2\gamma\omega\sin(2\omega t)}{4 M\beta \omega^2 \omega_0^2}
\end{equation}
which asymptotically reduces to the equilibrium variance as expected.

To make further progress with the non-Gaussian noise contribution, we focus in the sequel on Poissonian noise
\begin{equation}\label{poisson}
\mwe{\eta(u_k)\dots\eta(u_1)}=\lambda\alpha^k\delta(u_1-u_2)\dots\delta(u_{k-1}-u_k) \ , \ k>1\, .
\end{equation}
In an electrical circuit noise of this characteristics is approximately generated by a voltage-driven tunnel contact with a voltage much larger than $k_{\rm B} T$. Then, $\alpha$ corresponds to the transferred charge  and $\lambda$ is the transport rate. The cumulants in position for the harmonic oscillators follow from the known generating functional of the Poisson process
\begin{align}
{\mathcal F}_{\eta}[w]&=\lambda\int\limits_{-\infty}^{\infty}\dD u\left(\E^{\I\alpha w(u)}-1\right).\label{klass-rech_poissonvert}
\end{align}
according to ${\mathcal F}_{q_\eta}[w]={\mathcal F}_{\eta}[\tilde{w}]$.

\section{Classical Correlation Functions in Position Space}\label{sec:classical}

With the generating functional at hand we can now analyze the lowest order cumulants in more detail. The second cumulant is trivial as it is identical to $C_{q_\xi}$ up to a constant factor, namely, $C_{q_\eta}(t,t')=\lambda \alpha^2/(2M\gamma/\beta) C_{q_\xi}(t,t')$. Thus, the presence of the Poissonian noise leads to an additional heating and it can simply be taken into account by introducing an effective temperature $T_{\rm eff}=T+\lambda \alpha^2/(2M\gamma k_{\rm B})$.

More interesting is the three-point-correlation function
\begin{align}
	C_{q_\eta}(t,t,t)&=\frac{\lambda\alpha^3}{M^3\omega^3}\left\{\frac{\omega(4\omega_0^2-\gamma^2)}{6 \omega_0^2(2\gamma^2+\omega_0^2)}+\frac{\E^{-\frac{3}{2}\gamma t}}{24\omega_0^2}\right.\nonumber\\
&\left.\times\Bigg[2\omega\cos(3\omega t)+\gamma\sin(3\omega t)\right.\nonumber\\
&\left.-9\omega_0^2\frac{2\omega\cos(\omega t)+3\gamma\sin(\omega t)}{2\gamma^2+\omega_0^2}\Bigg]\right\}\, \label{eq:klass-rech_3moment-stationaer}
\end{align}
which yields asymptotically the stationary expression $C_{q_\eta}^{(3)}=\lim_{t\to \infty}C_{q_\eta}(t,t,t)$ with
\begin{equation}
C_{q_\eta}^{(3)}= \frac{2\lambda\alpha^3}{3M^3\omega_0^2(2\gamma^2+\omega_0^2)}\, .\label{eq:klass-rech_3moment-stationaer2}
\end{equation}
Hence, in contrast to $C_{q_\eta}^{(2)}$, the skewness becomes basically independent of friction for weak dissipation.

All higher order cumulants can be calculated accordingly. After exploiting (\ref{poisson}) the general expression reduces to
\begin{eqnarray}
&&C_{q_\eta}(t_N,\ldots, t_1)=\frac{\lambda \alpha^N\E^{-\frac{\gamma}{2}\sum_{k=1}^N t_k}}{M^N\omega^N}\nonumber\\
&& \times\sum_{m=0}^N \sum_{\Pi_{1N}}\prod_{i=1}^{N-m}\sin(\omega t_{p_i}) \prod_{j=1}^m\cos(\omega t_{p_{N-m+j}})\nonumber\\
&&\times(-1)^m \int\limits_{0}^{t_1}\dD u\, \E^{\frac{N\gamma}{2}u}\sin^m(\omega u)\cos^{N-m}(\omega u),
\end{eqnarray}
where in the second sum the indices $p_i$ run over all permutations $\Pi_{1N}$ of the set $\{1,\ldots, N\}$.
While no compact expression for the remaining integral is known, it can easily be calculated for specific values of $N, m$.

The stationary cumulants $C_{q_\eta}^{(N)}=\lim_{t\to \infty}C_{q_\eta}(t,\ldots, t)$ determine the non-Gaussianity of the steady state distribution in position of the oscillator.
As an example we give the results for some of them
\begin{eqnarray}\label{highercumu}
C_{q_\eta}^{(4)}&=& \frac{3 \lambda \alpha^4}{4 M^4 \omega_0^2\gamma (3 \gamma^2 + 4 \omega_0^2)}\nonumber\\
C_{q_\eta}^{(5)}&=& \frac{24 \lambda \alpha^5}{5 M^5\omega_0^2 (24\gamma^4+58 \gamma^2\omega_0^2+9\omega_0^4)}\nonumber\\
C_{q_\eta}^{(6)}&=& \frac{5 \lambda \alpha^6}{3M^6\omega_0^2\gamma (10 \gamma^4+37\gamma^2\omega_0^2+16\omega_0^4)}\, .
\end{eqnarray}
As already seen for the second and the third order cumulants above, the even order cumulants depend sensitively on the friction strength ($\propto 1/\gamma$) while the odd cumulants saturate in the weak dissipation regime. Note, however, that friction must be always finite to reach the steady state, otherwise the oscillator will continuously heat up. Hence, for $\gamma\ll \omega_0$ the even higher order cumulants always dominate against the odd ones, whereas in the overdamped limit $\gamma\gg \omega_0$ all cumulants are suppressed with $C_{q_\eta}^{(N)}\propto 1/\gamma^{(N-1)}$. As a consequence, the symmetry of the force field acts like a filter such as to favor even order cumulants with the tendency to establish a symmetric non-Gaussian steady state distribution.
For instance,  the relative weight of the non-Gaussian cumulants with respect to the Gaussian one $g_n=[{C}_{q_\eta}^{(n)}/{C}_{q_\eta}^{(2)} ] (M\omega_0/\alpha)^{n-2}$ reads for weak friction: $g_3=(\gamma/\omega_0)/6, g_4=3/8, g_5=(\gamma/\omega_0) 48/45, g_6=5/24$.

\section{Free Particle with Drude Damping}\label{sec:drude}

The above findings for a harmonic system can also be used to describe the dynamics of a free Brownian particle subject to Drude damping and driven by Poissonian noise. In this case, the damping kernel is $\gamma(t)=\gamma\omega_D\exp(-\omega_D t)$ which in the limit of the Drude frequency $\omega_D\to \infty$ reduces to the white noise situation. The most interesting regime is that of small $\omega_D$ though. Namely,
the Langevin equation
\begin{equation}\label{drude1}
\ddot{q}(t)+\int\limits_0^t\dD s\gamma(t-s)\dot{q}(s)=[\xi(t)+\eta(t)]/M
\end{equation}
can be mapped to that of an harmonic oscillator by putting $z(t)=\int_0^t\dD s\gamma(t-s)\dot{q}(s)$. One finds
\begin{equation}
\ddot{z}+\omega_D\dot{z}+\omega_D \gamma z =\gamma\omega_D [\xi(t)+\eta(t)]/M
\end{equation}
with $\dot{q}(t)=\dot{q}(0)-\int_0^t ds\, [M z(s)+\xi(s)+\eta(s)]/M$. Now, the Drude frequency takes the role of an effective friction constant, while $\gamma$ corresponds to the bare frequency of the fictitious $z$-oscillator. Accordingly, as in (\ref{inhomo-harmonic}) the inhomogeneous part of the solution is determined by the response function (\ref{eq:harm-oszi_greensfunktion}) with $\gamma\to \omega_D$ and $\omega\to \tilde{\omega}\equiv\sqrt{\omega_D\gamma-\omega_D^2/4}$.

\begin{figure}[htb]
  \centering
    \includegraphics[width=0.48\textwidth]{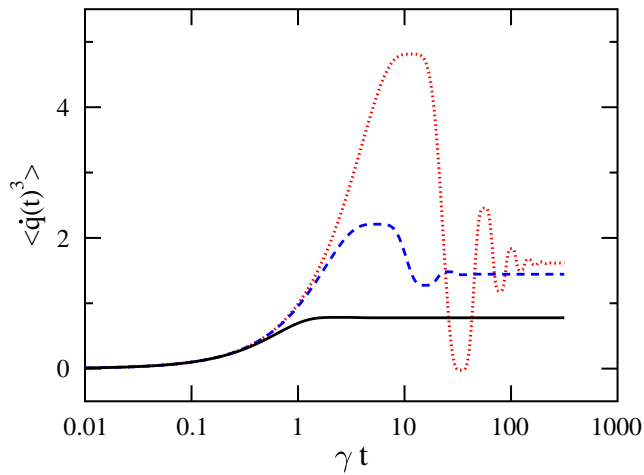}
      \caption{Dynamics of the third moment $\langle\dot{q}(t)^3\rangle$ in units of $\lambda\alpha^3/\gamma$ of a free particle immersed in a heat bath with Drude damping and subject to Poissonian noise in units of $\gamma$. The Drude frequency decreases from bottom to top: $\omega_D/\gamma=1$ (black, solid), $\omega_D/\gamma=0.1$ (blue, dashed), $\omega_D/\gamma=0.02$ (red, dotted).} \label{fig:q3}	
\end{figure}

For free Brownian motion the most interesting quantity is the velocity $\dot{q}(t)$ and its cumulants which can now be inferred from the findings of the previous sections. The resulting expressions are rather complex, but approach a stationary state after an oscillatory transient period (cf.~Fig.~\ref{fig:q3}). While for larger $\omega_D/\gamma$ the third moment $\langle \dot{q}(t)^3\rangle$ quickly saturates, large amplitude oscillations appear on a long-lasting transient period of time for small Drude frequencies. Correspondingly, the $z$-oscillator turns from an overdamped to an underdamped motion. In the stationary limit one finds
\begin{eqnarray}\label{drudecorr}
\lim\limits_{t\to\infty}\langle\dot{q}(t)^2\rangle &=& (2M\gamma k_{\rm B} T+\lambda \alpha^2)\frac{\gamma+\omega_D}{2\gamma\omega_D}\nonumber\\
\lim\limits_{t\to\infty}\langle{\dot{q}(t)^3}\rangle&=&\lambda \alpha^3 \frac{5\gamma+2\omega_D}{3\gamma^2+6\gamma\omega_D}\, .
\end{eqnarray}
We note in passing that $\langle \dot{q}(t)^2\rangle$ displays a much smoother transient behavior compared to $\langle \dot{q}(t)^3\rangle$  in the low friction regime (small $\omega_D/\gamma$).

A realization of this model is given by a voltage-biased tunnel junction in series with an impedance consisting of an $RC$-circuit (Ohmic resistance $R$ + capacitor $C$) and subject to Poissonian noise.  In this case, the phase across the junction is related to the voltage via $\phi(t)=(e/\hbar)\int_0^t ds V(s)$ and deviations $\tilde{\phi}=\phi-e V t/\hbar$ correspond to the generalized coordinate $q$ in (\ref{drude1}) \cite{ingold}. Its conjugate variable is the charge $\tilde{Q}$. The Drude frequency is identical to the inverse of the $RC$-time of the impedance.
  Hence, the velocity correlations (\ref{drudecorr}) specify charge fluctuations at the junction due to the Poissonian noise. According to Fig.~\ref{fig:q3} large charge oscillations occur on long time scales for high impedances (large $RC$-time).

\section{Quantum mechanics in coordinate space}\label{sec:qm_pos}

In the remainder we will generalize the results of  Sec.~\ref{sec:classical} to the quantum regime, first in coordinate space and then in energy space representation. In coordinate space the most straightforward way is to follow the lines described in \cite{grabert,weiss} as long as we are mainly interested in steady state expressions. The linearity of Heisenberg's equation of motion for the coordinate operator $\hat{q}(t)$ means that the quantum response function coincides with the classical one. Accordingly, the driving by Poissonian noise is taken into account exactly as in (\ref{inhomo-harmonic}). As a consequence, the non-Gaussian cumulants are identical to the results specified in (\ref{eq:klass-rech_3moment-stationaer2}) and (\ref{highercumu}). Only the second cumulant carries quantum mechanical information via the damping dependent equilibrium variance due to a heat bath with ohmic spectral density, i.e.,
\begin{equation}\label{equivariance}
\frac{\langle q^2\rangle_\beta}{q_0^2}=\frac{1}{\theta}+\frac{2}{\pi(\lambda_+-\lambda_-)}\left[\Psi\left(1+\frac{\theta \lambda_+}{4\pi}\right)-
\Psi\left(1+\frac{\theta\lambda_-}{4\pi}\right)\right]\, ,
\end{equation}
where $\theta=\omega_0\hbar\beta$ and $\lambda_\pm=(\gamma/\omega_0)\pm \sqrt{(\gamma/\omega_0)^2-4}$. Further the bare ground state variance is $q_0^2=\hbar/M\Omega_0$ and $\Psi$ denotes the di-Gamma function.

\begin{figure}[htb]
  \centering  	
    \includegraphics[width=0.48\textwidth]{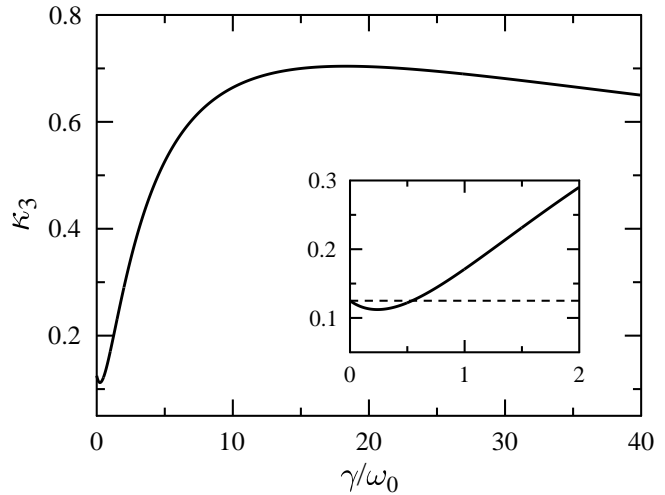}
     \caption{Dimensionless factor $\kappa_3$ the size of which determines the magnitude of the asymmetry in the coordinate distribution $\rho_3(q)$ vs.\ friction at zero temperature (see text for details). The inset displays a blow-up for weak damping with the dashed line corresponding to the undamped value $\kappa_3(\gamma/\omega_0=0)$.}\label{fig:kappa3}
\end{figure}

The reduced density operator of a harmonic system is obtained exactly within the path integral formalism by tracing out the Gaussian heat bath (for details see \cite{weiss}) and then averaging over the noise drive, i.e., $\rho(t)=\langle{\rm Tr}_R\{W(t)\}\rangle_\eta$ where $W(t)$ is the density operator of the full compound (system+reservoir). In coordinate representation $\rho(q,q')=\langle q|\rho|q'\rangle$ the steady state takes the form
\begin{eqnarray}
\label{quantum1}
\rho_{\rm st}(x,r) &\equiv&\lim_{t\to\infty}\langle r+\frac{x}{2}|\rho(t)|r-\frac{x}{2}\rangle\nonumber\\
&=&\frac{1}{\sqrt{2\pi \langle q^2\rangle_\beta}} \langle {\rm e}^{-\frac{[r-q_\eta(t\to \infty)]^2}{2\langle q^2\rangle_\beta}}\rangle_\eta\  {\rm e}^{-\frac{\langle p^2\rangle_\beta x^2}{2\hbar^2}}
\end{eqnarray}
with $r=(q+q')/2$, $x=(q-q')$ and $q_\eta$ as in (\ref{inhomo-harmonic}). The momentum variance $\langle p^2\rangle_\beta$ reads as given in \cite{grabert,weiss}, its specific form is not of relevance for the following discussion. The above distribution is no longer Gaussian with respect to the mean coordinate $r$ but rather attains an asymmetry and long tails due to the Poissonian noise drive. However, it is still normalized according to the bare Gaussian normalization factor. For a weak noise amplitude $\alpha$, an expansion of the diagonal part yields
\begin{eqnarray}\label{position-expansion}
\rho_{\rm st}(x,r)&=&\rho_\beta(x,r)\Bigg[1+\frac{r^2-\langle q^2\rangle_\beta }{2\langle q^2\rangle_\beta^2} \, C_{q_\eta}^{(2)}\nonumber\\
&&+r\, \frac{r^2-3\langle q^2\rangle_\beta}{6\langle q^2\rangle_\beta^3}\, C_{q_\eta}^{(3)}+O(\alpha^4)\Bigg]
\end{eqnarray}
with the classical cumulants $C_{q_\eta}^{(n)}$ and the bare thermal equilibrium $\rho_\beta(x,r)$ [results from (\ref{quantum1}) by putting $q_\eta=0$]. The asymmetry of the coordinate distribution $\rho_3(q)=[\rho_{\rm st}(0,q)-\rho_{\rm st}(0,-q)]/\rho_\beta(0,q)$ can now be probed to retrieve in leading order the skewness (third order cumulant) of the noise drive. Its magnitude depends crucially on the dimensionless factor $\kappa_3=(\langle q^2\rangle_\beta^3/q_0^6) (1+2\gamma^2/\omega_0^2)$ in the nominator of $\rho_3$ such that at a fixed coordinate a small $\kappa_3$ yields a large $\rho_3$. According to (\ref{equivariance}), for fixed damping $\kappa_3$ decreases with decreasing temperature. Particularly intriguing is thus the limit of zero temperature, where one has
\begin{equation}\label{q2}
\frac{\langle q^2\rangle_\infty}{q_0^2}=\left\{
\begin{array}{l l}
\frac{\omega_0}{2\omega}\left[1-\frac{2}{\pi}{\rm Arctan}\left(\frac{\gamma}{2\omega}\right)\right]\ & \ \gamma/\omega_0\leq 2\\
 & \\
\frac{2}{\pi (\lambda_+-\lambda_-)}\,  {\rm ln}(\lambda_+/\lambda_-) \ & \ \gamma/\omega_0\geq 2
\end{array}\right.
\end{equation}
with $\omega=\sqrt{\omega_0^2-\gamma^2/4}$ as above. In the domain of weak friction, this leads to $\kappa_3\approx 1/8 -3\gamma/(8\pi\omega_0)$, while for very strong friction $\kappa_3\approx 16\omega_0[\ln(\gamma/\omega_0)]^3/(\pi^3\gamma)$ (cf.~Fig.~\ref{fig:kappa3}). Hence, $\rho_3(q)$ is most pronounced for low damping at $\gamma/\omega_0\approx 0.25$ where $\kappa_3$ attains a minimum, or in the overdamped limit where  $\kappa_3$ decreases with increasing $\gamma/\omega_0$. Physically, friction always suppresses quantum fluctuations in  position [see (\ref{q2})] leading to a narrower Gaussian portion of the distribution. In contrast,
$C_{q_\eta}^{(3)}$ grows with increasing dissipation. The combination of these two factors gives rise to a minimum value of $\kappa_3$ for weak friction (the cumulant is almost independent of friction) and a decreasing value towards very strong friction (position fluctuations are strongly reduced).

 Recent experiments with current biased Josephson junctions as threshold detectors \cite{jj} operate in the classical regime. Then, for moderate to large friction the contribution of the third cumulant to the escape rate  follows up to a numerical factor from the asymmetry of the coordinate (superconducting phase) distribution in the harmonic well of a tilted washboard potential \cite{jjtheory}.   Our above results may provide signatures for the impact of quantum fluctuations when formally evaluating the contribution of the skewness at the position of the barrier top. Since for fixed friction, $\kappa_3$ decreases with decreasing temperature the relative impact of Poissonian noise is  enhanced due to quantum effects. This strongly indicates the advantage for operating in the low temperature domain which, however, necessitates  a consistent quantum mechanical description of the full anharmonic problem.

\section{Quantum mechanics in energy space}\label{sec:qm_energy}

In the previous section we have demonstrated the presence of non-Gaussian cumulants in the steady state distribution in coordinate space already in third order. Here, we focus on corresponding properties in energy space. These could in principle be inferred from the former results, however, we want to put our analysis in a somewhat broader context and consider a full dynamical equation (master equation).

For this purpose, one follows the conventional perturbative approach for a system $H_{\rm S} $ weakly coupled to a heat bath $H_{\rm B}$. An additional 'reservoir'  $H_{\rm NG}$ is added which generates general non-Gaussian noise. A general formulation is then based on a Hamiltonian
\begin{equation}
H=H_{\rm{S}}+H_{\rm{B}}+H_{\rm{I, B}}+H_{\rm{I, NG}}+H_{\rm{NG}}.
\end{equation}
The interaction is bilinear in the reservoir and system degrees of freedom, i.e.,
\begin{equation}
H_{\textrm{I, B}}+H_{\textrm{I, NG}}=g_{\rm B} Q \sum_j c_j x_j+g_{\rm NG} R\, M\, ,
\end{equation}
 where $Q$ and $R$ are system operators, the thermal bath $H_{\text{B}}$ is modeled as a collection of harmonic oscillator degrees of freedom $x_j$, and $M$ denotes an operator of the non-Gaussian reservoir. We do not need to specify the   non-Gaussian reservoir $H_{\text{NG}}$ (which in general may be very   complicated) since only its correlation functions enter the formulation   as we will see below. The reservoirs are statistically independent with $\langle \sum_j c_j x_j\rangle_{\rm B}=\langle M\rangle_{\rm NG}=\langle M \sum_j c_j x_j \rangle_{\rm B, NG}=0$.

The goal is now to derive an approximate equation of motion for the reduced density $\rho(t)$ within the Born-Markov perturbation theory up to third order in the coupling constants $g_{\rm B}$ and $g_{\rm NG}$ \cite{brosco2006}. After a straightforward calculation and using the eigenstates of the system $H_{\rm S}|n\rangle=E_n|n\rangle$, one obtains
in the interaction picture
\begin{align}
&\bra{m}\dot{\rho}_{\textrm{S,I}}(t)\ket{n}=\sum_{k,l}\bra{l}\rho_{\textrm{S,I}}(t)\ket{k}\nonumber\\
&\times\left[ \frac{g_{\rm B}^2}{\hbar^2}R_{knml}^{\textrm{G}}+\frac{{g}_{\rm NG}^2}{\hbar^2}R_{knml}^{\textrm{NG}}-\frac{\I {g}_{\rm NG}^3}{\hbar^3} C_{knml}^{\textrm{NG}}\right]\E^{\I(\omega_{kn}+\omega_{ml})t} \label{eq:qm-rech_mastergl-3.Ordnung}
\end{align}
Here, the second order contributions contain conventional Redfield tensors in rotating wave approximation the explicit form of which is well-known in the literature \cite{bp}. The non-Gaussian reservoir gives rise to a third order term $C_{knml}^{\textrm{NG}}$ that is lengthy and not very illuminating. Important to note is though that  it carries system matrix elements of the form
\begin{equation}
\bra{n}(\hat{a}\pm\hat{a}^\dag)^3\ket{n'}=\sum_{l,m}\bra{n}\hat{a}\pm\hat{a}^\dag\ket{l}\bra{l}\hat{a}\pm
\hat{a}^\dag\ket{m}\bra{m}\hat{a}\pm\hat{a}^\dag\ket{n'}\ .\label{eq:qm-rech_symmetrie-ueberlegung}
\end{equation}
These matrix elements vanish if $n=n'$ which has direct consequences on the appearance of this non-Gaussian contribution in the steady state.

To see that in detail, we consider a two level system $H_{\rm S}=\hbar\omega\sigma_z$ with the interaction Hamiltonian $H_{\textrm{I}}=\sigma_x (\tilde{g}_{\rm NG}M +g_{\rm B} \sum_j c_j x_j)$. The  $\sigma_x, \sigma_z$ denote Pauli matrices where $|0\rangle, |1\rangle$ is the eigenbasis of $\sigma_z$. At very low temperatures this system approximates the harmonic oscillator sufficiently well. From (\ref{eq:qm-rech_mastergl-3.Ordnung}) the relevant equations of motion
in the Schr\"odinger picture  are obtained as
\begin{eqnarray}\label{master-equation}
\dot{\rho}_{00}&=&-(W_{10}+W_{01})\, \rho_{00}-2 \Lambda \rho_{01}'' + W_{10}\nonumber\\
\dot{\rho}_{01}''&=&2\Lambda {\rho}_{00}-\Lambda- 2 D' \rho_{01}'' +(2 D'' +\omega) \rho_{01}'\nonumber\\
\dot{\rho}_{01}'&=& -\omega {\rho}_{01}'\, ,
\end{eqnarray}
 where we exploited that $\rho_{11}+\rho_{00}=1$ and $\rho_{01}\equiv \rho_{01}'+i \rho_{01}''=\rho_{10}^*$. The conventional transition rates read
\begin{equation}
W_{10/01}=\int ds \left[\frac{g_{\rm B}^2}{\hbar^2} \langle {\mathcal E}(s) {\mathcal E}(0)\rangle +\frac{g_{\rm NG}^2}{\hbar^2} \langle M(s) M(0)\rangle\right] {\rm e}^{\pm i\omega s}
\end{equation}
with ${\mathcal E}=\sum_j c_j x_j$. The other second order contribution $D=D'+ i D''$ introduces a Lamb-shift and is given by
\begin{equation}
D=\int ds \left[\frac{g_{\rm B}^2}{\hbar^2} \langle \{{\mathcal E}(s), {\mathcal E}(0)\}\rangle +\frac{g_{\rm NG}^2}{\hbar^2} \langle \{M(s), M(0)\}\rangle\right] {\rm e}^{i\omega s}
\end{equation}
 with $\{,\}$ being the anti-commutator. In (\ref{master-equation}) the leading non-Gaussian contribution appears  as the real-valued parameter
\begin{equation}
\Lambda =\frac{2 g_{\rm NG}^3}{\hbar^3} \int_0^\infty ds \int_0^s du \langle M(s) M(u) M(0) + M(0) M(s) M(u)\rangle\, .
\end{equation}
It is derived from the tensor $C_{knml}^{\textrm{NG}}$ by assuming noise frequencies to be much larger than the system frequency $\omega$ (white noise limit).

Given specific initial conditions, the above master equation can now easily be solved. For the ground state population the result is
\begin{align}
\rho_{00}(t)&=\frac{W_{10}}{\Gamma}+\E^{-\Gamma t}\left(\rho_{00}(0)-\frac{W_{10}}{\Gamma}\right)\nonumber\\
&\phantom{=}+\frac{4\Lambda\rho''_{01}(0)}{\Gamma}\E^{-\Gamma t}\left(\E^{-\frac{1}{2}\Gamma t}\cos\omega_{\text{R}}t -1\right)
\end{align}
using the abbreviations $\Gamma\equiv W_{10}+W_{01}$ and $\omega_{\text{R}}^2=2\omega D''+\omega^2-\Gamma^2/4$.
Obviously,  the  third moment influences only the transient dynamics but is absent in the steady state behavior which here coincides with the thermal distribution. The reason for that can be traced back to the structure (\ref{eq:qm-rech_symmetrie-ueberlegung}):  in steady state detailed balance for the populations dictates this non-Gaussian contribution to be absent.

This can also be seen from the result in coordinate space (\ref{eq:klass-rech_3moment-stationaer}) directly.
The energy representation follows from
\begin{eqnarray}
\rho_{{\rm st}, nm}&=&\langle n|\rho_{\rm st}|m\rangle\nonumber\\
&=& \int dq dq' \psi_n\left(r+\frac{x}{2}\right)\ \rho_{\rm st}(r,x)\, \psi_m\left(r-\frac{x}{2}\right)\,
\end{eqnarray}
with wave functions $\psi_n(q)=\langle n|q\rangle$ and eigenstates $|n\rangle$ of the harmonic system. Now, in the weak damping limit according to (\ref{eq:qm-rech_mastergl-3.Ordnung}) off-diagonal elements $\rho_{{\rm st}, n\neq m}$ are zero and one can concentrate on the populations $\rho_{{\rm st}, nn}$.
The part in (\ref{position-expansion}) originating from the  third cumulant is anti-symmetric in $r$, while the bare equilibrium $\rho_\beta(x,r)$ is symmetric with respect to $x, r \to -x, -r$. The same is true for the product of the two wave functions with $n=m$ so that the integral vanishes and the population is indeed independent of the third cumulant.

We note that in contrast to the finding of the weak damping limit, the result (\ref{position-expansion}) is valid for any coupling to the  thermal bath and only perturbative in the non-Gaussian noise. For stronger dissipation off-diagonal elements remain finite $\rho_{{\rm st}, n\neq m}\neq 0$ and the third cumulant appears in the steady state distribution.\\

\section{Conclusions}\label{sec:conclusions}
In conclusion we have investigated a linear detector (dissipative harmonic oscillator) in the presence of Gaussian and non-Gaussian noise in the classical and the quantum regime.
Starting from the classical response function of the problem and the cumulant-generating function we investigated position correlation functions beyond second order.  For the special case of Poissonian noise we found a general rule for the calculation of cumulants of arbitrary degree and gave an explicit expression for the three-point-correlation function.  While the even order cumulants depend sensitively on the damping strength, the odd order cumulants saturate in the weak dissipation regime. Consequently, for $\gamma\gg\omega_0$ odd higher order cumulants will always be dominated by the even ones, while in the limit $\gamma\ll\omega_0$ the damping suppresses all cumulants uniformly. Consequently, the harmonic potential tends to support a symmetric non-Gaussian steady state distribution.

These results are mathematically identical to the situation of a free particle with Drude damping, as e.g.\ realized in the case of a voltage-biased tunnel junction in series with an impedance consisting of an RC-circuit and subject to Poissonian noise. Its velocity distribution displays an interesting dependence on the circuit parameters. For $\omega_{\text{D}}\ll\gamma$ the skewness oscillates substantially before the system reaches equilibrium, whereas for $\omega_{\text{D}}\geq\gamma$ the transient to the steady state is monotonous.
 In the quantum regime but still with classical Poissonian noise, the position representation of the reduced density operator of the harmonic system was obtained exactly within the path integral formalism. It turns out that due to the damping induced suppression of quantum fluctuations in position, the skewness of the distribution is most pronounced either for weak or for very strong friction.
 Results for general types of white non-Gaussian noise can be obtained within the conventional Born-Markov approximation. The corresponding master equation for a two level system reveals that in the energy representation  to lowest order non-Gaussian noise effects are absent in steady state distributions. Simple symmetry arguments can be found for the appearance of non-Gaussian noise in position representation.

Our findings help to understand observations in recent experiments with current biased Josephson junctions where the contribution of the third cumulant to the escape rate coincides up to a numerical factor with the asymmetry of the harmonic position distribution in the harmonic well of a tilted washboard potential. The quantum fluctuations predicted here  may become visible as these experiments leave the classical regime.

\section*{Acknowledgements}
We thank H. Grabert, J. Pekola, and H. Pothier for fruitful discussions. Financial support was provided by the DFG through SFB569 and the DAAD.

\end{document}